# A novel method to determine the layer number of 2D TMD materials based on Optical Microscopy and Image Processing

Bilal Bera Meriç[1], Ayşe Erol[1] and Fahrettin Sarcan[1*]

[1] Department of Physics, Faculty of Science, Istanbul University, Vezneciler, Istanbul 34134, Turkey.

*e-mail: fahrettin.sarcan@istanbul.edu.tr

**Abstract**

Two-dimensional (2D) transition metal dicakcoganite (TMD) materials have unique electronic and optical properties. The electronic band structures of the materials alter as a function of layer numbers, which results in modifications to the whole characteristic properties. Therefore, the determination of the layer number is crucial for optoelectronic applications. In this study, a fast and easily applicable method is proposed for determining the layer number of two-dimensional TMD materials by using an optical microscopy and computatial methods. The method uses image processing techniques on digital images taken with a Complementary Metal Oxide Semiconductor (CMOS) camera under a conventional reflecting microscope. The chromaticity differences and lightness difference in International Commission on Illumination (Commission internationale de l'éclairage ,CIE) Luv color space values between the layered areas on the flakes and substrates are used together to train the model to be identified the layer numbers of the materials and corrected via photoluminescence spectroscopy. The random forest clasifier algorithm is applied to predict layer numbers on unknown materials. This approach provides high accuracy under varying microscope configurations such as brightnesses, gain etc. and material-substrate combinations. From ML to bulk whole layers numbers of $MoS_2$, $WSe_2$ and $WS_2$ flakes are determined with high accuracy by training only a single flake of each has various layer numbers. Compared with traditional methods such as Raman spectroscopy, atomic force microscopy (AFM) and photoluminescence (PL), our method is not only faster but also easier to apply. Moreover, unlike the prominent methods in the literature such as machine learning or clustering-based, it requires only single training for a material/substrate combination and further accelerates the processes.

## 1. Introduction

Over the last two decades, two-dimensional (2D) materials have been highly demanded in various scientific and technological research and applications, because of their extraordinary optical, electronic, and structural properties [1–3] 2D material family is a large membered material family from insulator to conductors, there are a numerous different member such as Graphene (Gr) [4], black phosphorus (BP) [5], hexagonal boron nitride (h-BN) [6], metal nitrides/carbides (MXenes) [7], metal–organic frameworks (MOFs)[8] , a family of monoelemental compounds (Xenes)[9] , metal oxides, perovskite-type oxides[10], 2D polymers [11] and transition metal dichalcogenides (TMDs, $MX_2$)[12] etc.

The structural, optical and electrical properties of these materials not only change from material to material but can also significantly vary with the number of layers [13–16]. For example, a single-layer of Gr exhibits unique characteristics like massless Dirac fermions and high carrier mobility [17,18], whereas multilayer graphene may display poor electrical properties due to interlayer interactions [17,19]. Likewise, TMDs with bandgaps and strong light-matter interactions, which are ideal for optoelectronics and photovoltaics applications [20,21], and the number of layers in TMDs impact their bandgap, thereby allowing for tailoring optical and electronic properties [1,16,22,23]. Therefore, determination of the layer number of 2D materials is crucial due to the significant variations in their properties.

To determine the number of layers of 2D TMD materials, Raman spectroscopy, Atomic Force Microscopy (AFM) and photoluminescence spectroscopy are widely used in the literature [24]. All these non-destructive methods, though effective and provide reliable results, however can be time-consuming and labor-intensive. On the other hand, optical microscopy has become a quicker and accessible method for determining the thickness of 2D materials, thanks to advancements in image processing and machine learning techniques. There are a few studies in the literature use optical microscope images of 2D materials to determine the layer numbers. Wang et. al. developed a method using optical microscopy and image processing based on the optical contrast of the layers [25] . The contrast values were calculated using red, green and blue (RGB) channel values in optical images between substrate and material and they were able to determined the layers of Gr, $MoS_2$ flakes, and reported that these contrast values increased linearly with material thickness. Sterbentz et al. developed an image segmentation program using unsupervised clustering algorithms for the automatic thickness identification of 2D materials from digital optical microscop images [26]. They analysed all three digital colour channels and applied Gaussian mixture model fits to data clusters. The results demonstrated a pixel accuracy of approximately 95% in identifying mono- and few-layer flakes on both opaque and transparent substrates. Dong et. al. developed a method using deep learning for microscopic imagery classification, segmentation, and detection [27] .Three deep-learning architectures (DenseNet, U-Net, and Mask-RCNN) were benchmarked on tasks involving multi-label classification, segmentation, and detection of 2D material TMDs. They reported that these models effectively identified mono-, bi-, tri-, multilayer, and bulk flakes using microscopic images of $MoS_2$ fabricated on $SiO_2$/Si substrates. However, in optical contrast-based methods, the accuracy of the results obtained may vary depending on the light conditions and optical microscope settings. Futuremore, the methods based on RGB channel analysis can be limited to only certain substrate and material combinations. Different substrates or environmental light changes can affect the consistency of contrast values, limiting the generalizability of the method. Although the image segmentation methods which uses clustering algorithms provides significant advantages in terms of automatic analysis, preliminary adjustments appropriate to the distribution of the data must be made in order for the clustering algorithms to provide optimal results. This may require additional parameter settings and testing in different 2D flakes or different optical conditions. Additionally, the computational costs and processing time of the model used may increase in large data sets. On the other hand, the deep learning-based approach require intensive resources in terms of time and processing power in the data preparation and model training stages, as it requires a large amount of labeled data for model training. Although the high accuracy of deep learning architectures is seen as an advantage, the high processing power requirement during the training and testing process may limit the practical applications of these methods.

In this study, we introduce a rapid and accurate single-sample training method based on chromaticity and lightness differences, utilizing Random Forest classification, to compare color relationships layer numbers.This approach provides a faster and more straightforward alternative to traditional machine learning techniques such as k-means clustering and k-nearest neighbors (k-NN). Our approach, unlike methods relying on iterative optimization algorithms, requires minimal computational resources and is less sensitive to parameter tuning. Furthermore, compared to other machine learning methods like hierarchical clustering or Gaussian mixture models, our proposed method simplifies the segmentation process by utilizing pre-segmented reference samples, thereby reducing the need for complex training procedures and computational overhead. By using the Luv color space, we aim to more accurately calculate the color and brightness differences between the material and the substrate. For this purpose, we train a random forest classification model with parameters obtained from chromaticity and lightness differences. To simplify the process, only one masked and labeled reference image is required for each material and substrate pair. The materials employed in the study include $MoS_2$, $WS_2$, and $WSe_2$. A detailed theoretical framework is presented under the title "Theoretical Background" in order to explain the basis of these methods and the principles on which they are based.

## 2. Theoritical Background

For low-dimensional materials, an increase in thickness (or the number of layers for the layered materials) causes more light to be reflected due to interference effects, leading to changes in the intensity and spectrum of the reflected light. This phenomenon can be explained using the Fresnel equations, which describe the reflection and transmission of electromagnetic waves at the material. The reflectivity coefficient $R$ for unpolarized light at normal incidence is given by:

$$R = \frac{\left(n_0^2+n_2^2\right) sin^2\left(\frac{2\pi n_1 d}{\lambda}\right)}{\left(n_0^2+n_2^2\right) sin^2\left(\frac{2\pi n_1 d}{\lambda}\right)+4n_0^2 n_2^2} \qquad (1)$$

where $n_0$, $n_1$, and $n_2$ are the refractive indices of the surrounding medium, the thin film, and the substrate, respectively; $d$ is the thickness of the the film, and $\lambda$ is the wavelength of incident light. As the number of layers increases in layered materials, the optical interference effects become more pronounced, influencing the reflected light's intensity.

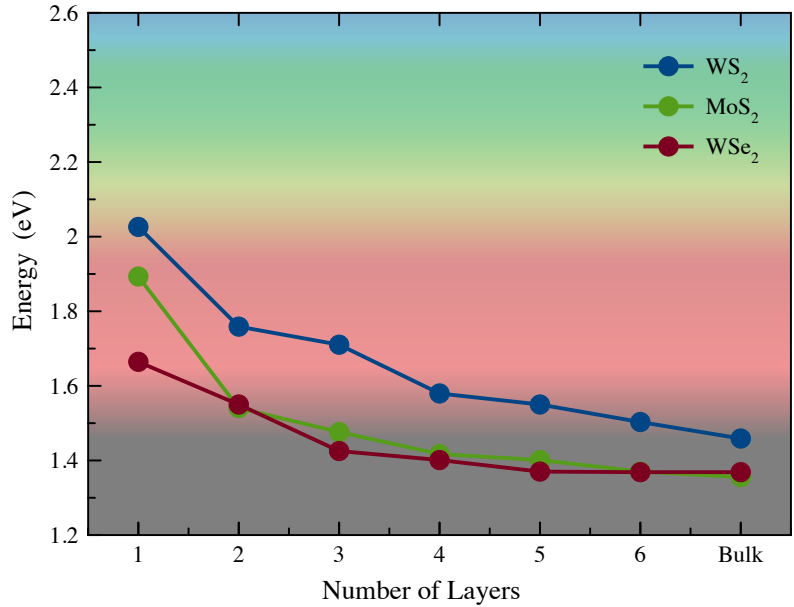

**Figure 1.** Experimentaly determined bandgap values of some 2D TMDs as a function of layer number

The number of layers also alters the band structure of 2D materials, which in turn affects their optical and electrical properties. For example, many 2D TMDs exhibit a larger bandgap in monolayer form, but bandgap value decreases as the number of layers increases (Figure 1.). This change is primarily due to dependency of electronic band structure of the materials to the layer number and quantum confinement effects, which is less significant in thicker materials. The changes in the band gap affects the wavelength of absorbed and reflected light. 2D TMD materials with narrow band gaps absorb light at longer wavelengths, which also affects the spectrum of reflected light.

## 3. Experimental and Computational Methods

### 3.1 Sample Preparation

Bulk single-crystal TMD materials were purchased from 2Dsemiconductorâ and hpGrapheneâ companies. A scotch tape and polydimethylsiloxane (PDMS)-assisted mechanical exfoliation method was used to obtain 2D TMD flakes with different numbers of layers [13] .The flakes with different number of layers were transferred onto the $SiO_2$ substrates by using a custom-built viscoelastic transfer system.

### 3.2 Optical Characterization

To carry out optical characterisations, a conventional reflecting optical microscope and micro PL spectroscopy were used. The PL spectroscopywere used as a complementary tool in the training process of the model. The optical microscope uses a tungsten halogen lamp, and it does not contain any optical filters. The PL spectrum of the flakes were taken using a custom-built modular micro spectroscopy setup equipped with a spectrometer (Shamrock 500i, Andor) and a thermoelectric cooled Si CCD (Newton BEX2-DD, Andor) and PylonIR InGaAs CCD. It is equipped with a 532 nm CW laser (Gem532, Novanta Photonics) as an excitation laser with ∼0.8 $\mu m$ beam diameter (with 100X, NIR, 0.7 NA objective) [13, 24].

### 3.3 Random Forest Classification

Random Forest is one of the ensemble learning methods and consists of the combination of multiple decision trees. Each decision tree is trained independently on a randomly selected subset of the training data. This process is known as "bagging" (bootstrap aggregation) and increases generalizability by reducing the risk of overfitting the model. In each decision tree, only a randomly selected subset of features is used for data splitting. This feature bagging increases the diversity and independence between the trees, thus increasing the accuracy of the model. For the classification problem, Random Forest combines the class prediction of each tree with the majority voting method. Each tree makes a class prediction and the class with the most votes is accepted as the final prediction. This allows the model to produce more robust and accurate results. In addition, Random Forest offers a "feature importance" calculation method that helps determine which features are more effective in the model decisions. In this way, the features that contribute the most to the accuracy of the model can be determined and feature selection can be made [32].

## 4. Results and Discussion

### 4.1 Training Process

The model training process basically consists of several steps (Figure 2). The training process is based on the processing of the optical image of the sample and the analysis of the chromaticity and lightness differences obtained from this image and the training a random forest clasification model.

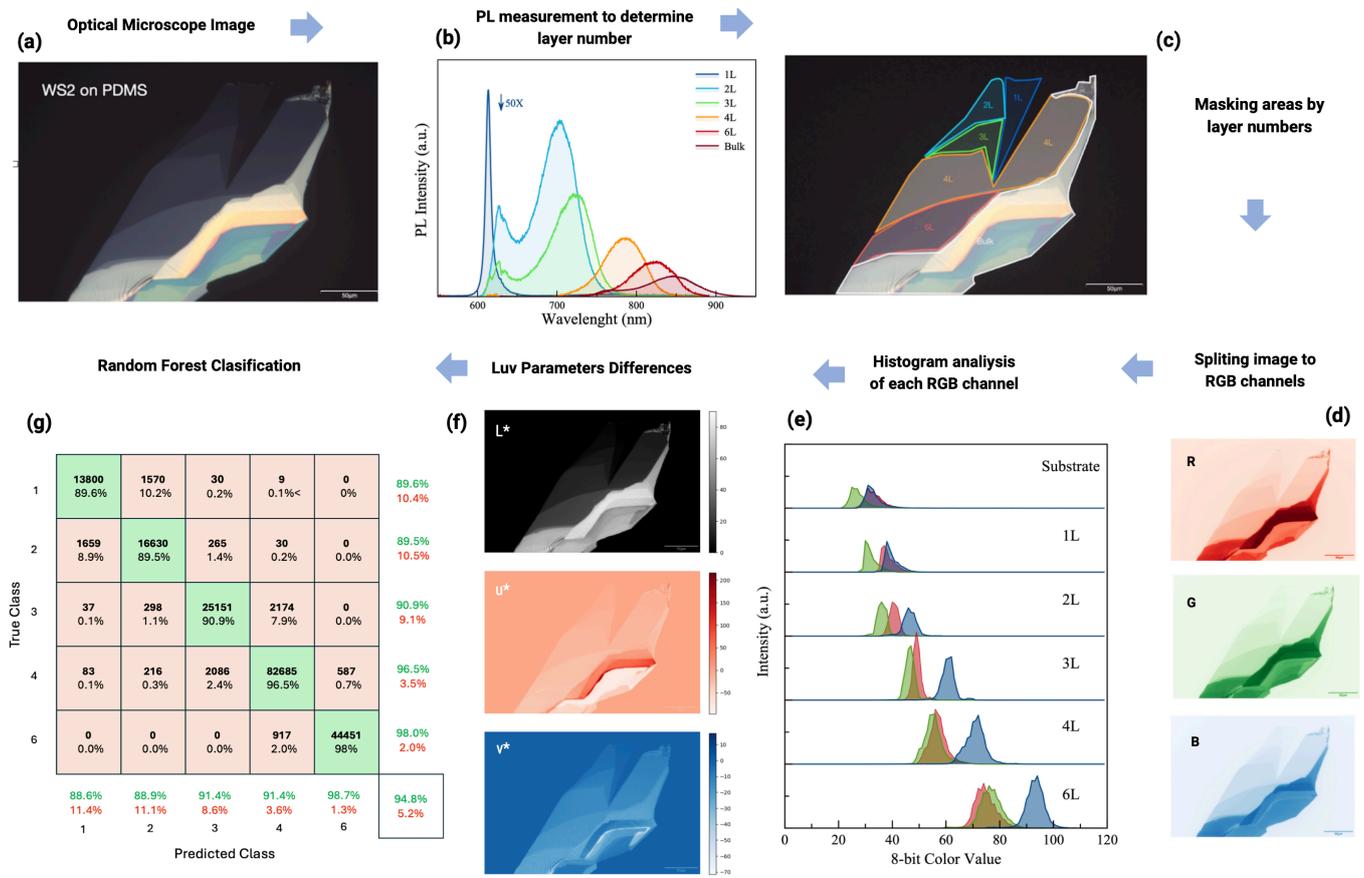

**Figure 2.** (a) Optical microscopy image of $WS_2$ flake on PDMS substrate, (b) PL spectrums of the regions to verify layer numbers. (c) masked image by layer, (d) heatmaps maps for image separated into each RGB channel, (e) Color histogram of RGB channels for layered areas and substrate, (f) heatmaps for $\Delta L$, $\Delta u$ and $\Delta v$ parameters between layered layered areas and subtrate and (g) confusion matrix of random forest classification model.

### 4.1.1 Imaging, Identifying and Masking

As the first step, the mechanically exfoliated WS₂ sample onto the PDMS substrate examined under an optical microscope. The sample illuminated with a halogen tungsten lamp and a image captured with a CMOS camera at 40X objective (Figure 2a). In the model training process, a single flake image with different number of layer is sufficient to classify the regions in the number of layers of the flake.

The number of layers on the material were determined by PL measurements (Figure 2b) and the layer information is then used for the masking process. Each area is masked with different colours; for example, a single layer is masked in red and two layers are masked in blue. Unidentified regions or bulk regions are masked in white, and substrate regions are not included in the masking process (Figure 2c).

**4.1.2 Calculating The Chromaticity and Lightness Differences**

In the next stage, the pixels in the masked regions were separated, and the lightness and chromaticity differences between the layered regions and the substrate were analyzed. This process began by extracting the color profile of the substrate and reducing it to a single RGB color value (2d). For this, a histogram was created for each R, G, and B value. In digital images, an RGB color histogram is a graph that shows the pixel intensity distribution of each of the red, green, and blue color components in an image. This histogram is obtained by counting how many pixels are present at different intensity levels for each color component in the image. Each color component is represented by intensity values ranging from 0 to 255, which are segmented into 256 bins in the histogram.

In this study we utilize the standard method for calculating RGB histograms, which involves splitting an image into its red, green, and blue channels and counting pixel intensities for each channel. Mathematically, the histogram $H_c(i)$ for channel c at intensity level i is expressed as:

$$H_c(i) = \sum_{(x,y) \in \text{Image}} \delta(I_c(x,y) - i)) \qquad (2)$$

where $I_c(x, y)$ is intensity of channel c at pixel (x,y) and δ is the Dirac delta function, defined as:

$$\delta(z) = \begin{cases} 1 & \text{if } z = 0, \\ 0 & \text{otherwise.} \end{cases} \qquad (3)$$

The histogram represents the frequency distribution of 8-bit color values (0–255 for each channel) across the pixels. The frequency values on the histograms are normalized between 0 and 1. The shift in the color channels by the layer number can be see on the Figure 2e.

Using the histogram, the weighted mean for the color values was calculated, with the frequency values serving as the weights. The weighted mean was computed using the following equation:

$$\bar{X} = \frac{\sum_{i=1}^{n} W_i X_i}{\sum_{i=1}^{n} W_i} \qquad (4)$$

where the $W_i$ represent the intensities on the histogram and $X_i$ is the 8-bit colour values. By employing this calculation, we reduced the entire pixel area of substrate area into a single colour value. By employing this calculation, the entire pixel area of the substrate was reduced to a single color value, which was then used for further analysis of lightness and chromaticity differences.

Afterwards, before convert colors to the L*u*v* color space the First, RGB values are linearized because regular CMOS cameras apply gamma correction to colors [31]. This process made the brightness and contrast of the images independent of the camera, thus reflecting the actual optical properties. Linearization was performed using the following equation:

$$C_{\text{linear}} = \begin{cases} \dfrac{C_{RGB}}{12.92} & \text{if } C_{RGB} \leq 0.04045 \\ \left(\dfrac{C_{RGB} + 0.055}{1.055}\right)^{2.4} & \text{if } C_{RGB} > 0.04045 \end{cases} \qquad (5)$$

where the C component represents any of the RGB values (R, G, or B).

The CIE $L^*u^*v^*$ col r space was developed by the International Commission on Illumination (Commission internationale de l'éclairage ,CIE) in 1976, specifically to analyse the colour differences of light sources better [28,29] . The LUV colour space defines colours with three components: $L^*$ represents the lightness value, and $u^*$ and $v^*$ both express the hue and saturation in different wavelength range, respectively [29,30] . This system helps to perceive colour differences more accurately under different lighting conditions [28,29] . The LUV color space is designed to represent color differences in a way that is closer to how the human eye perceives colors, and aims to provide a uniform perception which means that color changes are perceptually equal in all color tones [28,30] .

$L^*$ varies between black (0) and white (100), indicating how bright or dark the color is. The $u^*$ and $v^*$ parameters are chromaticity components. $u^*$ represents the red-green axis, while $v^*$ denotes the blue-yellow axis. Negative $u^*$ values indicate greener tones, while positive $u^*$ values signify reddish tones. Similarly, negative $v^*$ values correspond to bluish tones, whereas positive $v^*$ values represent yellowish tones.

Before converting linearized RGB values to the CIE L*u*v* color space, we first need to convert RGB values to the CIE 1931 XYZ color space [29]. The CIE 1931 XYZ color space is a device-independent standard that represents colors consistently. A key feature of this color space is its perceptual homogeneity, meaning that changes in perceived colors correspond to equal changes in XYZ values [32,33].

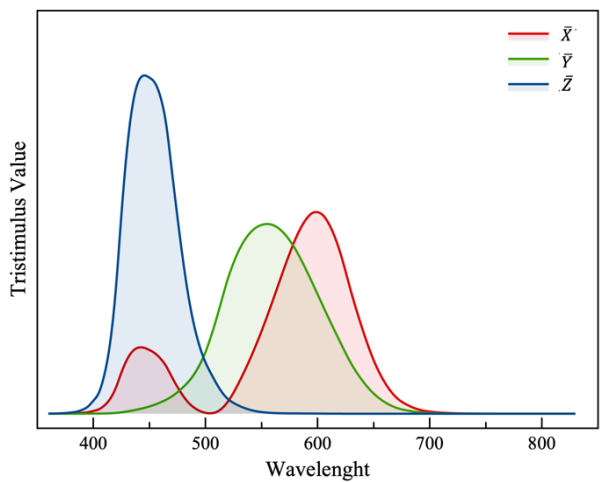

**Figure 3.** CIE color-matching functions (CIE 1964 10° observer)

Furthermore, switching to the XYZ color space when converting from RGB to CIE L*u*v* provides a standardized and perceptually consistent representation, minimizing distortions and ensuring more accurate color transformations[29,33] . The CIE 1931 XYZ color matching functions (Figure 3) represent the human eye's response to different wavelengths and form the foundation for defining the XYZ color space [28,29]. Converting from RGB to XYZ relies on a transformation matrix derived from the RGB primaries [32] .

An important factor to consider during this conversion is the selection of the illuminant. Different illuminants vary according to the light source, its spectral power distribution (SPD), and how its energy is distributed across the spectrum [28,30]. In this study, a tungsten halogen lamp was used as the light source on an optical microscope. According to CIE standards, the light from a tungsten lamp is defined as Illuminant A [28] which has a warmer and reddish spectrum. [28,30]. For comparison, Illuminant D65 represents a standard daylight illuminant with a cooler and bluish spectrum, commonly used in colorimetric studies. Similarly, power LEDs offer a highly customizable spectral distribution and are increasingly employed in applications requiring specific wavelength peaks. Figure 4 shows the spectral power distributions (SPDs) of different light sources, including D65, a power LED, and Illuminant A.

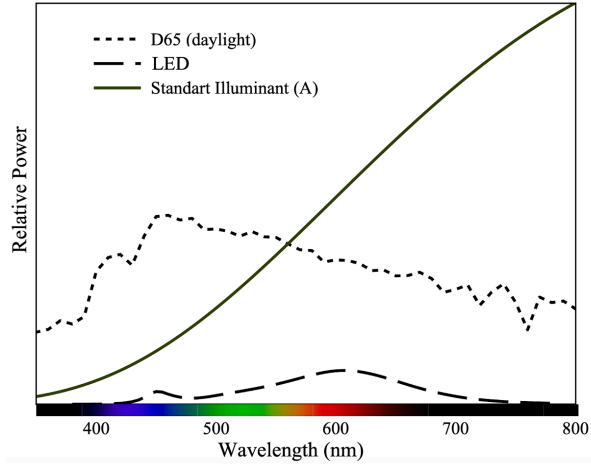

**Figure 4.** Spectral power distribution of D65, power LED and halogen lamp.

The following equations used to convert linearized RGB values to CIE 1931 XYZ colour space.

$$X = k \int S(\lambda)\bar{x}(\lambda)d\lambda \quad Y = k \int S(\lambda)\bar{y}(\lambda)d\lambda \quad Z = k \int S(\lambda)\bar{z}(\lambda)d\lambda \quad (6)$$

$S(\lambda)$ represents the spectral power distribution of the light source (Illuminant A). $\bar{x}(\lambda)$, $\bar{y}(\lambda)$ and $\bar{z}(\lambda)$ are the CIE colour matching functions [28,29]. To convert RGB values directly to XYZ from RGB values, these integrals can be reduced to conversation matrix.

$$\begin{bmatrix} X \\ Y \\ Z \end{bmatrix} = \begin{bmatrix} R_{11} & R_{12} & R_{13} \\ R_{21} & R_{22} & R_{23} \\ R_{31} & R_{32} & R_{33} \end{bmatrix} \cdot \begin{bmatrix} R_{linear} \\ G_{linear} \\ B_{linear} \end{bmatrix} \quad (7)$$

Each matrix element represents the contribution of the RGB primaries to the XYZ tristimulus values, depending on the color space and illuminant [32]. The reference white point for Illuminant A is defined as $(X_r, Y_r, Z_r)=(1.09850, 1.00000, 0.35585)$ [28]. The parameters for CIE L*u*v* colour space from XYZ values were calculated using following equations;

$$L^* = f(x) = \begin{cases} 116\sqrt[3]{y_r} - 16, & if\ y_r > 0.008856 \\ 903.3 y_r, & otherwise \end{cases} \quad (8a)$$

$$u^* = 13L(u' - u'_r) \qquad v^* = 13L(v' - v'_r)$$

where,

$$u' = \frac{4X}{X+15Y+3Z} \quad (8b)$$

$$v' = \frac{9Y}{X+15Y+3Z} \quad (8c)$$

$$u'_r = \frac{4X_r}{X_r+15Y_r+3Z_r} \quad (8d)$$

$$v'_r = \frac{9Y_r}{X_r+15Y_r+3Z_r} \quad (8e)$$

here, $Y_r$ is the luminance of the reference white for Illuminant A.

The CIE L*u*v* color space is employed in this study due to its perceptual uniformity, ensuring accurate and meaningful analysis of color differences. We calculated the L*, u*, and v* parameter differences between the layered areas and the substrate color (Figure 2e). These parameters are calculated separately to capture variations in lightness (L*) and chromaticity (u* and v*) along the red-green and blue-yellow axes. This approach allows us to observe how layer thickness influences color properties, providing a detailed understanding of the optical changes in the material.

**4.1.3 Dataset Creation and Model Training**

The L*, u* and v* differences for each layered region used as independent variables (features) in the dataset. This process was repeated for each layered region to obtain a large dataset. As a result, a data point for each layer number was represented by the L* difference, u* difference and v* difference values. These differences numerically define the color and brightness differences observed between the substrate and the layer. After calculating these differences for each layer, separate features were created for different regions such as 1 layer, 2 layer and 3 layer so on. Then, each data point in the dataset was labelled according to the number of layers which has been corrected by their PL spectrums. For example, when the differences were calculated for a 1 layered region, this data point was labelled as "1 layer", 2 layer regions were labelled as "2 layers" and 3 layered regions were labelled as "3 layers". These labels used as dependent variables for the model to estimate which regions had how many layers while performing classification.

Using this dataset, we trained the Random Forest classification model with the following hyperparameters: n_estimators=200, max_depth=20, min samples split=5, min samples leaf = 2, max features=0.75, bootstrap=True, and random state= 42. These parameters were chosen to prevent overfitting of the model and to increase the classification accuracy. In this way, we were able to capture more complex optical differences between the layers and the substrate, thus increasing the classification accuracy. The model training and prediction processes were conducted on a MacBook Air (2023 model) equipped with an Apple M2 processor (8-core CPU, 10-core GPU) and 16 GB of unified memory.

**4.1.4 Reliability and Accuracy Test**

To evaluate the performance of the Random Forest model, the model was tested with data obtained from different examples of the images used in the training process. In this way, the generalization ability of the model was evaluated and its capacity to correctly predict the number of layers in different materials were tested. The classification performance of the model was analysed with the confusion matrix (Figure 2g). The matrix shows which layer numbers where the model classified correctly and in which cases it made an error.

**4.2 Tests and Results**

To test the ability of the method to determine the number of layers in different materials, we used it on $WS_2$, $MoS_2$ and $WSe_2$ samples. Polydimethylsiloxane (PDMS) substrate was chosen for each material. PDMS allows 2D TMD materials to be exfoliated smoothly onto the surface and enables the layer thickness to be determined as it is exfoliated without requiring a transfer process. To evaluate the sensitivity of the method to environmental factors, images were obtained by different researchers using different CMOS cameras and microscopes. The shots were taken at various exposure times and illumination intensities. The models trained for each material gave highly consistent results after they trained separately for each materials. The overall accuracies of the model and the importance levels of the ΔL, Δu, and Δv parameters for each material are presented in Table 1. These results show the sensitivity of the model to the optical properties of each material and which parameters are more effective in layer number estimation.

**Table 1.** Model accuracy values and feature importances of parameters for the tested materials.

| Material | Model Accuracy | $\Delta L^*$ Importance | $\Delta u^*$ Importance | $\Delta v*$ Importance |
|---|---|---|---|---|
| **$WS_2$** | %95 | %43 | %31 | %26 |
| **$MoS_2$** | %94 | %43 | %28 | %29 |
| **$WSe_2$** | %95 | %72 | %17 | %11 |

### 4.2.1 Testing on WS$_2$

We trained the model by using image in Figure 2a it has all layers from 1 to 6 except 5 layer. According to the model results, WS$_2$ is a material where especially L* and color-based parameters (u* and v*) are important due to the strong relationship between bandgap change and layer thickness. While monolayer WS$_2$ has a larger bandgap, the bandgap decreases as the number of layers increases, and this affects both the intensity and wavelength of the reflected light. This change increased the sensitivity of the model to L* and color parameters. In particular, WS$_2$ exhibits strong color changes in the visible range (Figure 1), which caused color-based parameters to have higher importance in the model. However, in thicker flakes, the bandgap energy level stabilizes and the effect of this parameter becomes weaker. The confusion matrix of the model can be seen on Figure 2g. We tested model on useen WS$_2$ images, the results can be seen on Figure 5.

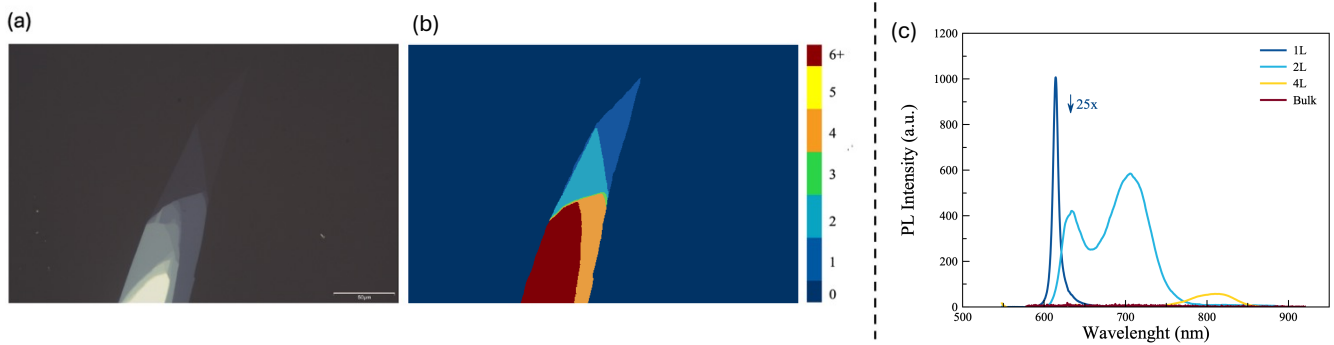

**Figure 5.** (a) Optical microscopy image of WS$_2$ on a PDMS substrate, (b) the predicted layer number of WS$_2$ and (c) photoluminescence (PL) result of the WS$_2$ flake to verify the predicted layer number.

### 4.2.2 Testing on MoS$_2$

To train the model for MoS$_2$ we used 2 images have all layers 1 to 6 except 5 layer. MoS$_2$ is a material that shows changes in bandgap energy level depending on the number of layers in the visible region (Figure 1). Especially in the transition between monolayer and several layers, the bandgap change is quite pronounced, which increases the importance of u* and v* parameters in the model. In thick layers, the bandgap change no longer occurs in the visible region. However, this large bandgap change in the transition from monolayer to higher layers causes shifts in the wavelength of the reflected light and strengthens the effect of color-based parameters. While the bandgap change becomes limited in thicker layers, it partially reduces the effect of color parameters, while increasing the importance of L* parameter in the model. This shows that both color and reflection properties are used in a balanced way in the segmentation for MoS$_2$. The confusion matrix of the model can be seen on Figure S1a. We tested model on MoS$_2$ images not incluted to the training process, the results can be seen on Figure 6.

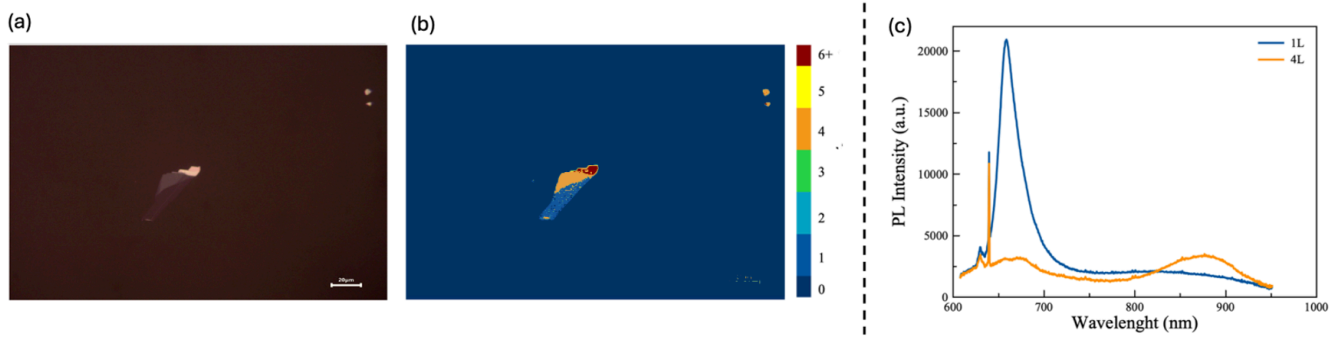

**Figure 6.** (a) Optical microscopy image of MoS$_2$ on a PDMS substrate, (b) the predicted layer number of MoS$_2$ and (c) photoluminescence (PL) result of the MoS$_2$ flake to verify the predicted layer number.

### 4.2.3 Testing on WSe$_2$

A single image used to train model for WSe$_2$ has all layers from 1 to 6 layers. WSe$_2$ is a material where the bandgap changes depending on the number of layers are quite limited and generally do not show any significant change in the visible region.

This situation caused the L* parameter to play a more dominant role in the model. The limited bandgap change caused the shifts in the color of the reflected light (expressed by the u and v parameters) to be minimal, thus reducing the importance of the color parameters. It make importance of L* parameter in the model high. The fact that the bandgap change in thicker layers occurs almost completely outside the visible region shifted the sensitivity of the model to the optical properties more to the reflection intensity. The confusion matrix of the model can be seen on Figure S1b. We tested model to make prediction on $WS_2$ images, the results can be seen on Figure 7.

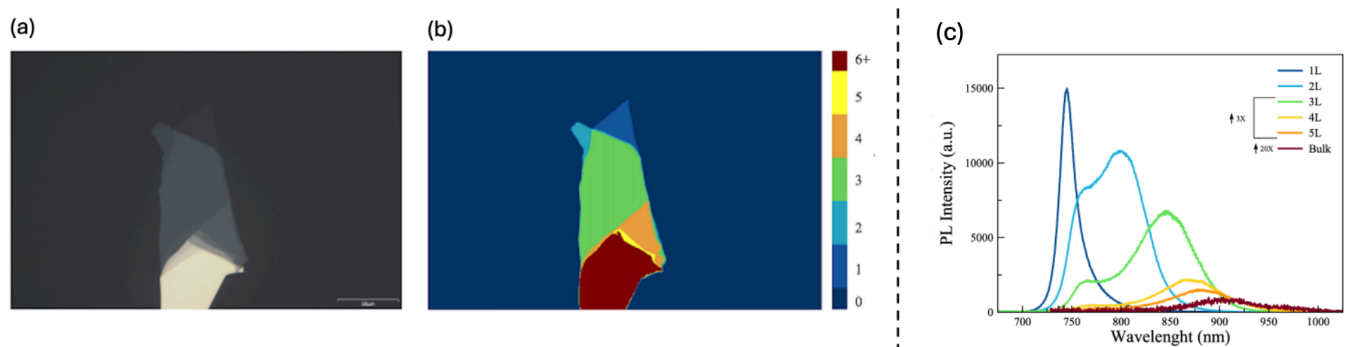

**Figure 7.** (a) Optical microscopy image of $WSe_2$ on a PDMS substrate, (b) the predicted layer number of $WSe_2$, and (c) photoluminescence (PL) of the $WSe_2$ flake to verify the predicted layer number.

## 5. Conclusion

This study presents a fast and efficient method for determining the layer numbers of two-dimensional transition metal dichalcogenides materials using optical microscopy and image processing techniques. The proposed method is based on analyzing the color and brightness differences between the material and the substrate, and uses a random forest classifier trained with data obtained from only a single reference image. This method offers a useful approach, especially after the exfoliation process, in terms of facilitating rapid prediction of layer number. The results show that the model achieves high accuracy across a variety of two-dimensional transition metal dichalcogenides materials and is a strong alternative to traditional methods.

The training process for the random forest classifier takes about a minute when a single reference image is used. The prediction process is extremely efficient and each prediction is completed in only 1 to 5 seconds depends ond the image size. These features reveal that the proposed method is a practical and applicable tool for quickly and easily determining the layer thickness of two-dimensional transition metal dichalcogenides materials. We tested our method on the more samples, the test results can be seen in supplementary information. The study demonstrates that the method is easily adaptable to different material and substrate combinations by demonstrating its ability to classify layer numbers with minimal training data. Future work could aim to extend this approach to a broader range of materials, enhance its robustness under varying environmental conditions, and improve dynamic prediction capabilities during the examination of two-dimensional transition metal dichalcogenides flakes under the microscope following the exfoliation process.

## DATA AVAILABILITY

The authors declare that all the datasets supporting the findings of this study are available from the corresponding authors on reasonable request.

## AUTHOR CONTRIBUTIONS

B.B.M. and F.S. fabricated the samples and captured optical microscope images. F.S. performed the steady-state photoluminescence measurements. B.B.M. and F.S. developed the model, and B.B.M. designed the computational framework. B.B.M., A.E., and F.S. analyzed the data. B.B.M. wrote the manuscript with contributions from all authors. A.E. and F.S. managed various aspects of the project and provided funding.

## Acknowledgements

F.S. gratefully acknowledges the support from the Scientific Research Projects Coordination Unit of Istanbul University (FBA-2023-39412) and The Scientific and Technological Research Council of Turkey (TUBITAK) project (121F169).

**Supplementary informations:** A novel method to determine the layer number of 2D TMD materials based on Optical Microscopy and Image Processing

Bilal Bera Meriç[1], Ayşe Erol[1] and Fahrettin Sarcan[1*]

[1] Department of Physics, Faculty of Science, Istanbul University, Vezneciler, Istanbul 34134, Turkey.

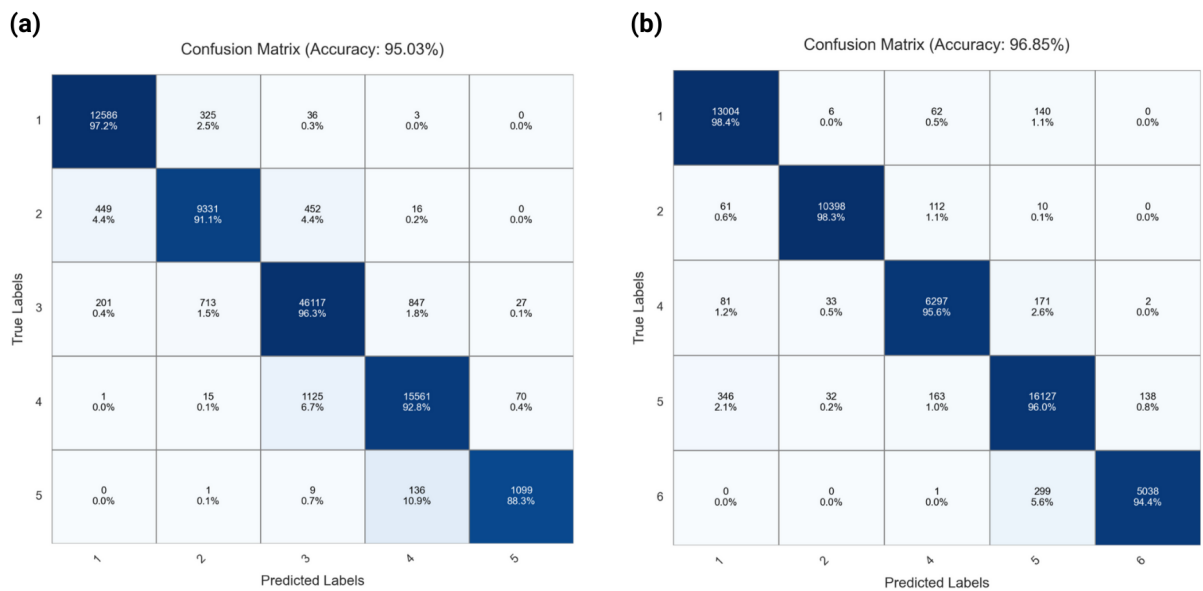

**Figure S1.** (a) Confusion matrix of the Random Forest model for $WSe_2$, illustrating the classification performance with true versus predicted labels. (b) Confusion matrix of the Random Forest model for $MoS_2$, providing a comparative view of model accuracy and misclassifications for this material.

**Table S1.** Combined performance metrics of the Random Forest model for $MoS_2$, $WS_2$, and $WSe_2$. The table reports precision, recall, F1-score, and support for different classification classes. Macro Average represents the unweighted mean of all classes, while Weighted Average accounts for class distribution.

| Material | Class | Precision | Recall | F1-Score | Support |
|---|---|---|---|---|---|
| $MoS_2$ | 1 | 0.96 | 0.98 | 0.97 | 13212 |
| | 2 | 0.99 | 0.98 | 0.99 | 10581 |
| | 4 | 0.95 | 0.96 | 0.95 | 6584 |
| | 5 | 0.96 | 0.96 | 0.96 | 16806 |
| | 6 | 0.97 | 0.94 | 0.96 | 5338 |
| | Macro Avg. | 0.97 | 0.97 | 0.97 | - |
| | Weighted Avg. | 0.97 | 0.97 | 0.97 | - |
| $WS_2$ | 1 | 0.88 | 0.90 | 0.89 | 15309 |
| | 2 | 0.89 | 0.89 | 0.89 | 18584 |
| | 3 | 0.91 | 0.91 | 0.91 | 27821 |
| | 4 | 0.96 | 0.97 | 0.96 | 85657 |
| | 6 | 0.99 | 0.98 | 0.98 | 45368 |
| | Macro Avg. | 0.93 | 0.93 | 0.93 | - |
| | Weighted Avg. | 0.95 | 0.95 | 0.95 | - |
| $WSe_2$ | 1 | 0.95 | 0.97 | 0.96 | 12950 |
| | 2 | 0.90 | 0.91 | 0.99 | 10248 |
| | 3 | 0.97 | 0.96 | 0.96 | 47905 |
| | 4 | 0.94 | 0.93 | 0.93 | 16772 |
| | 5 | 0.92 | 0.88 | 0.96 | 1245 |
| | Macro Avg. | 0.93 | 0.93 | 0.90 | - |
| | Weighted Avg. | 0.95 | 0.95 | 0.97 | - |

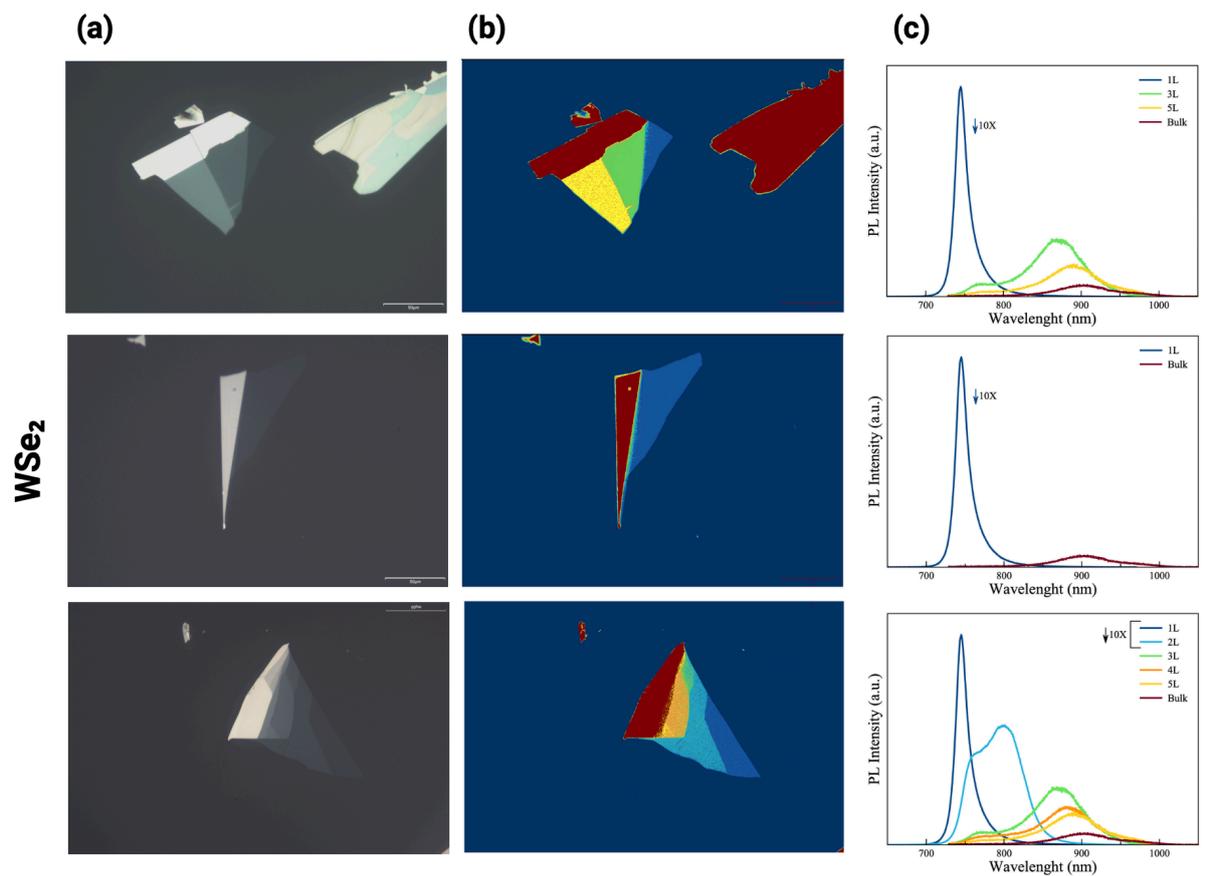

**Figure S2.** (a) Optical microscope images of WSe$_2$ samples. (b) Predicted layer numbers of samples. (c) PL results to verify the predictions.

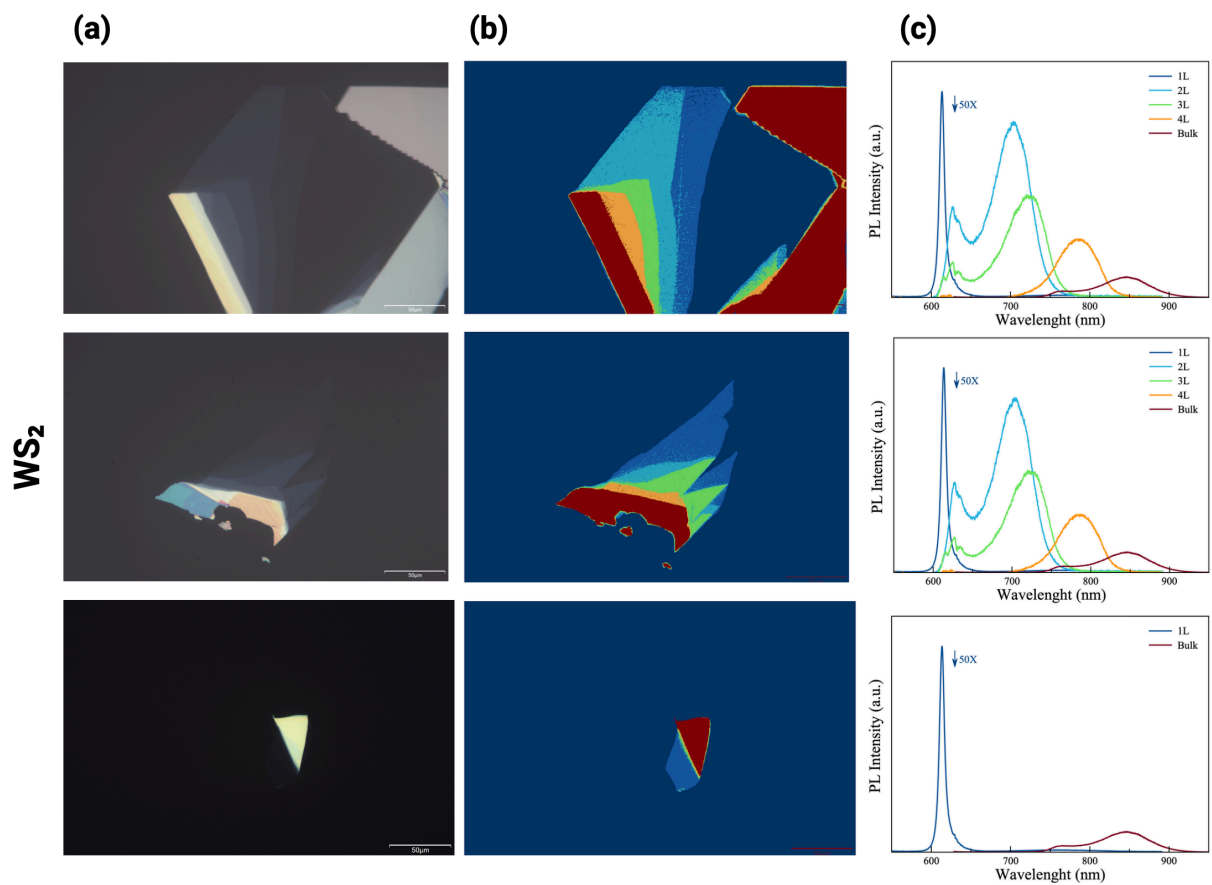

**Figure S3.** (a) Optical microscope images of WS$_2$ samples. (b) Predicted layer numbers of samples. (c) PL results to verify the predictions.

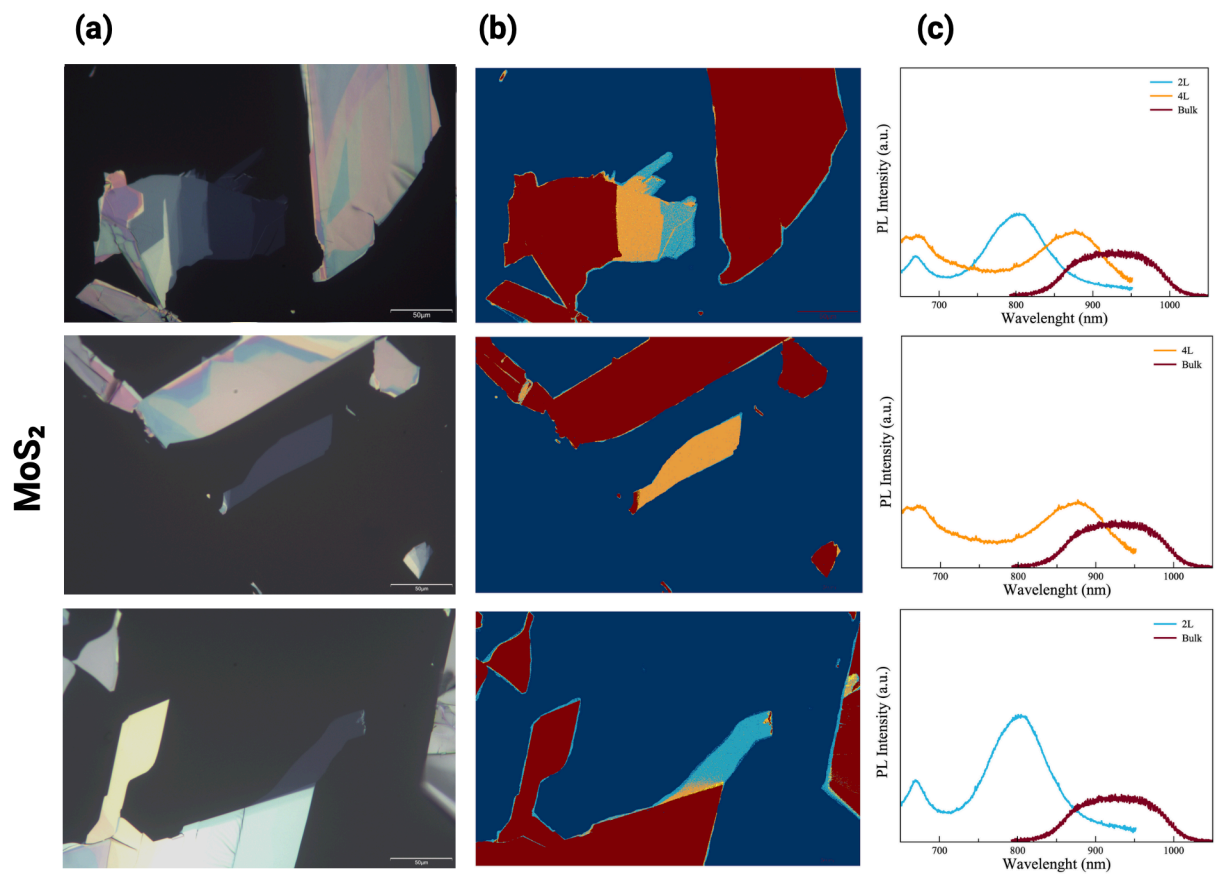

**Figure S4** (a) Optical microscope images of MoS$_2$ samples. (b) Predicted layer numbers of samples . (c) PL results to verify the predictions.

**The full Python code is availible at https://github.com/berameric/2d_material_layer_number_prediction**